\documentclass[aps, prapplied, superscriptaddress, reprint]{revtex4-2}

\usepackage{graphicx}
\usepackage{physics}

\usepackage[dvipsnames]{xcolor}

\newcommand{\um}[0]{\mu\mathrm{m}}

\newcommand{\fSAW}[0]{f_\mathrm{SAW}}
\newcommand{\lSAW}[0]{\lambda_\mathrm{SAW}}

\newcommand{\uij}[0]{u_{ij}}
\newcommand{\uxx}[0]{u_{xx}}
\newcommand{\uyy}[0]{u_{yy}}
\newcommand{\uxy}[0]{u_{xy}}

\newcommand{\vB}[0]{\mathbf{B}}
\newcommand{\B}[0]{B}

\newcommand{\vS}[0]{\mathbf{S}}
\newcommand{\Dm}[0]{\Delta m_S}
\newcommand{\mS}[0]{m_S}
\newcommand{\vbMW}[0]{\mathbf{b}_\mathrm{MW}}

\newcommand{\VSi}[0]{\mathrm{V_{Si}}}

\begin{document}


\title{Identification of acoustically induced spin resonances of Si vacancy centers in 4H-SiC} 

\author{Thomas Vasselon}
\affiliation{Universit\'e Grenoble Alpes, CNRS, Grenoble INP, Institut N\'eel, 38000 Grenoble, France}
\author{A.~Hern\'{a}ndez-M\'{i}nguez}
\email{alberto.h.minguez@pdi-berlin.de}
\affiliation{Paul-Drude-Institut f\"{u}r Festk\"{o}rperelektronik, Leibniz-Institut im Forschungsverbund Berlin e.V.,\\ Hausvogteiplatz 5-7, 10117 Berlin, Germany}
\author{M.~Hollenbach}
\affiliation{Helmholtz-Zentrum Dresden-Rossendorf, Institute of Ion Beam Physics and Materials Research, Bautzner Landstrasse 400, 01328 Dresden, Germany}
\affiliation{Technische Universit\"{a}t Dresden, 01062 Dresden, Germany}
\author{G.~V.~Astakhov}
\affiliation{Helmholtz-Zentrum Dresden-Rossendorf, Institute of Ion Beam Physics and Materials Research, Bautzner Landstrasse 400, 01328 Dresden, Germany}
\author{P.~V.~Santos}
\affiliation{Paul-Drude-Institut f\"{u}r Festk\"{o}rperelektronik, Leibniz-Institut im Forschungsverbund Berlin e.V.,\\ Hausvogteiplatz 5-7, 10117 Berlin, Germany}


\date{\today}

\begin{abstract}
The long-lived and optically addressable spin states of silicon vacancies ($\VSi$) in 4H-SiC make them promising qubits for quantum communication and sensing. These color centers can be created in both the hexagonal ($V1$) and in the cubic ($V2$) local crystallographic environments of the 4H-SiC host. While the spin of the $V2$ center can be efficiently manipulated by optically detected magnetic resonance at room temperature, spin control of the $V1$ centers above cryogenic temperatures has so far remained elusive. Here, we show that the dynamic strain of surface acoustic waves can overcome this limitation and efficiently excite magnetic resonances of $V1$ centers up to room temperature. Based on the width and temperature dependence of the acoustically induced spin resonances of the $V1$ centers, we attribute them to transitions between spin sublevels in the excited state. The acoustic spin control of both kinds of $\VSi$ centers in their excited states opens new ways for applications in quantum technologies based on spin-optomechanics. 
\end{abstract}


\maketitle 

\section{Introduction}

Atom-like color centers in SiC are attractive systems for applications in quantum technologies~\cite{Castelletto:2013jj, Kraus:2013vf, Koehl:2015kw, Radulaski:2017ic, Awschalom:2018ic, Atature:2018hh}. The most prominent example is the negatively charged silicon vacancy ($\VSi$)~\cite{Janzen:2009ij}. This center emits in the near infrared range, where optical glass fibers have low absorption, and has long-living spin states, which can be optically addressed and controlled by microwave (MW) fields~\cite{Koehl:2011fv, Baranov:2011ib, Riedel:2012jq}. Silicon vacancies in the 4H-SiC polytype can occupy two non-equivalent sites with hexagonal ($h$) and cubic ($k$) local crystallographic environments. This difference leads to two types of $\VSi$ centers, labeled as $V1$ and $V2$, respectively~\cite{Ivady:2017bq}, as shown in Fig.~\ref{fig:intro}(a). Both types of centers share the same half-integer spin $S = 3/2$, but the transition frequencies between their orbital and spin energy levels have different values due to the non-common local environments~\cite{Ivady:2017bq}.

The $V2$ center has been intensively studied since its MW-induced spin transitions in the ground state can be efficently addressed by optically detected magnetic resonance (ODMR) even at room temperature~\cite{Soltamov:2012ey, Kraus:2013di, Widmann:2014ve, Carter:2015vc, Niethammer:2016bc, Simin:2017iw, Embley:2017bf}. Moreover, similar to an oscillating magnetic field, elastic vibrations can also induce room-temperature spin transitions in this center~\cite{HernandezMinguez:2020kv}. In contrast to MW-driven spin resonances, which only allow changes in spin number $\Dm=\pm1$~\cite{Tarasenko:2017ky}, acoustic fields can induce spin transitions with $\Dm=\pm1$ and $\Dm=\pm2$ both in the ground and excited state multiplets~\cite{HernandezMinguez:2021sa}.

The strong ODMR contrast (almost 100\%) of the $V1$ center under resonant optical excitation~\cite{Nagy:2018ey} enables the high fidelity reading of the spin state. The latter also makes this center a promising quantum system for the realization of robust spin-photon interfaces~\cite{Togan:2010hi, Soykal:2016tk}, although its spin coherence time is not as long as its $V2$ counterpart~\cite{Castelletto:2020cj}. However, MW-driven spin manipulation and optical detection have so far only been reported at cryogenic temperatures~\cite{Nagy:2018ey, Nagy:2019fw}, which severely limits the use of the $V1$ center in quantum communication and sensing protocols. In this manuscript, we demonstrate that acoustic vibrations in the form of surface acoustic waves (SAWs) can efficiently manipulate the spin states of both the $V1$ and $V2$ centers up to room temperature. Using a spectrally filtered ODMR technique, we reveal SAW-driven spin resonances that are specific to spin transitions of the $V1$ center. Based on their width and temperature dependence, we attribute these resonances to transitions between spin sublevels in the excited state.

The manuscript is organized as follows. In Section~\ref{sec:ExpDet} we describe the sample and measurement method, as well as the main characteristics of the spin system under our experimental conditions. Section~\ref{sec:Results} shows our experimental results, while Section~\ref{sec:Discussion} discusses the origin of the spin resonances attributed to the $V1$ center. Finally, in Section~\ref{sec:Conclusion} we summarize the results of the manuscript and suggest possible applications in them in quantum technology applications.

\section{Experimental Details}\label{sec:ExpDet}

\begin{figure}
\includegraphics[width=\columnwidth]{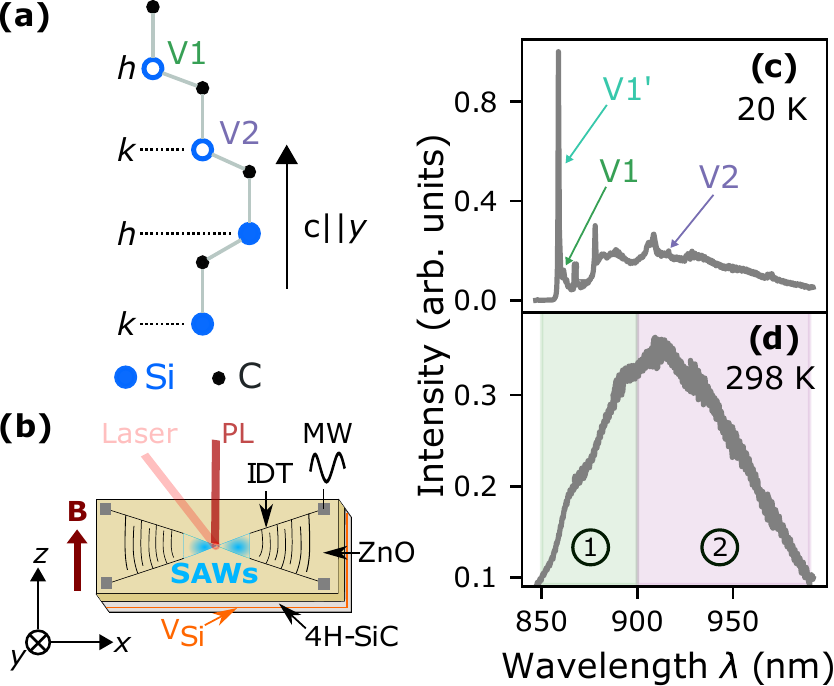}
\caption{\label{fig:intro} \textbf{(a)} Schema of the two non-equivalent sites for the silicon atom in 4H-SiC, which correspond to the $V1$ ($h$ site) and the $V2$ ($k$ site) centers. \textbf{(b)} Sketch of the hybrid spin-optomechanical system. It consists of a 4H-SiC wafer containing $\VSi$ centers at a well defined depth and coated with a ZnO thin film. An acoustic cavity consisting of two focusing IDTs is patterned on the ZnO film to excite SAWs. \textbf{(c)} Photoluminescence spectrum of the 4H-SiC at 20 K. \textbf{(d)} Same as (c), but measured at room temperature. The areas marked as 1 and 2 indicate the PL spectral regions used in the aODMR experiments.}
\end{figure}

Figure~\ref{fig:intro}(b) displays the hybrid spin-optomechanical system used in this work. It consists of a 4H-SiC substrate containing an ensemble of $\VSi$ centers created at a depth of $2.5~\um$ by proton irradiation with an energy of 375~keV and a fluence of $10^{15}$~cm$^{-2}$~\cite{Kraus:2017cka}. After irradiation, the SiC substrate was coated with a 35-nm-thick SiO$_2$ layer followed by a 700-nm-thick ZnO piezoelectric film using radio-frequency magnetron sputtering. Finally, acoustic cavities defined by a pair of focusing interdigital transducers (IDTs) were patterned on the surface of the ZnO film by electron beam lithography and lift-off metallization. Each IDT consists of 80 aluminum finger pairs for excitation/detection of SAWs with a wavelength $\lSAW=6~\um$ and a frequency $\fSAW\approx920$~MHz, and an additional Bragg reflector consisting of 40 finger pairs placed on its back side. The finger curvature and separation between the opposite IDTs ($\approx 120~\um$) are designed to focus the SAW beam at the center of the cavity.

The experiments were performed in a confocal microphotoluminescence ($\mu$PL) setup with the sample placed in a cold-finger cryostat equipped with a window for optical access and radio-frequency connections for the application of MW signals to the IDTs. The $\VSi$ centers were optically excited by a Ti-Sapphire laser at a wavelength of 780~nm focused onto a spot size of $10~\mu$m by a 20x objective with 0.4 numerical aperture. The PL from the centers was collected by the same objective and analyzed by a monochromator equipped with a charge-coupled device camera. Figure~\ref{fig:intro}(c) shows the PL spectrum of the $\VSi$ centers measured at 20~K. Three zero-phonon lines (ZPLs) are observed at 858~nm ($V1$'), 862~nm ($V1$) and 917~nm ($V2$) superimposed on their corresponding broad phonon sidebands (PSBs)~\cite{Nagy:2018ey}. The $V1$ and $V2$ lines are the optical transitions between the ground state (GS) and the first excited state (ES) of the centers, while the $V1$' line is the optical transition between the GS and the second ES of the $V1$ center~\cite{Janzen:2009ij}. The other lines in the spectrum are probably associated with features from the SiC substrate. At room temperature, the broad PSBs dominate the PL emission and the ZPLs are no longer observed, see Fig.~\ref{fig:intro}(d). 

The acoustically induced ODMR (aODMR) studies were performed by detecting the PL integrated over the phonon sidebands using a silicon photodiode. The laser stray light was removed from the optical path by a long pass dichroic mirror (805~nm), while a set of long and short pass filters selected the PL spectral range to be detected. To distinguish between the two types of centers, we exploit the fact that they emit in different wavelength ranges. For the aODMR measurements of the $V1$ centers, we collected the photons emitted in the 850-900~nm spectral range, see the green region marked as 1 in Fig.~\ref{fig:intro}(d) (note that it contains photons originating from the optical relaxation of both the first and second ES of the $V1$ center). For the $V2$ center, however, the photodiode detected the PL emitted above 900~nm, see the purple region marked as 2 in Fig.~\ref{fig:intro}(d). The SAWs were generated by applying an amplitude-modulated MW signal of appropriate frequency to one of the IDTs, and the output of the photodetector was connected to an amplifier locked-in to the MW modulation frequency.

The spin transition frequencies of the $\VSi$ centers are tuned to the SAW frequency by applying an in-plane magnetic field perpendicular to the SAW propagation direction, see~Fig.~\ref{fig:intro}(b). Here, we use a rotated reference frame where the $x$ axis is parallel to the SAW propagation direction, the $y$ axis points along the out-of-plane $c$ axis of the 4H-SiC, and $z$ is parallel to $\vB$. The effective spin Hamiltonian of the GS and first ES, expressed in the rotated reference frame and in the uniaxial approximation~\cite{Simin:2016cp}, read:

\begin{equation}\label{eq:SpinHamiltonian}
\mathcal{H}_{0}^{(GS,ES)} \; =  \; D^{(GS,ES)} \; ( S_{y}^{2} \; - \; \frac{5}{4}) \; + \; g \mu_B B S_{z} \;,
\end{equation}

\noindent where $g\approx2$ is the Landé g-factor, $\mu_B$ the Bohr magneton, $\vS=(S_x, S_y, S_z)$ the 3/2-spin operator with $S_z$ parallel to $\vB$, and $D^{(GS,ES)}$ are the zero field splitting (ZFS) constants.

\begin{figure}[ht!]
\includegraphics[width=\columnwidth]{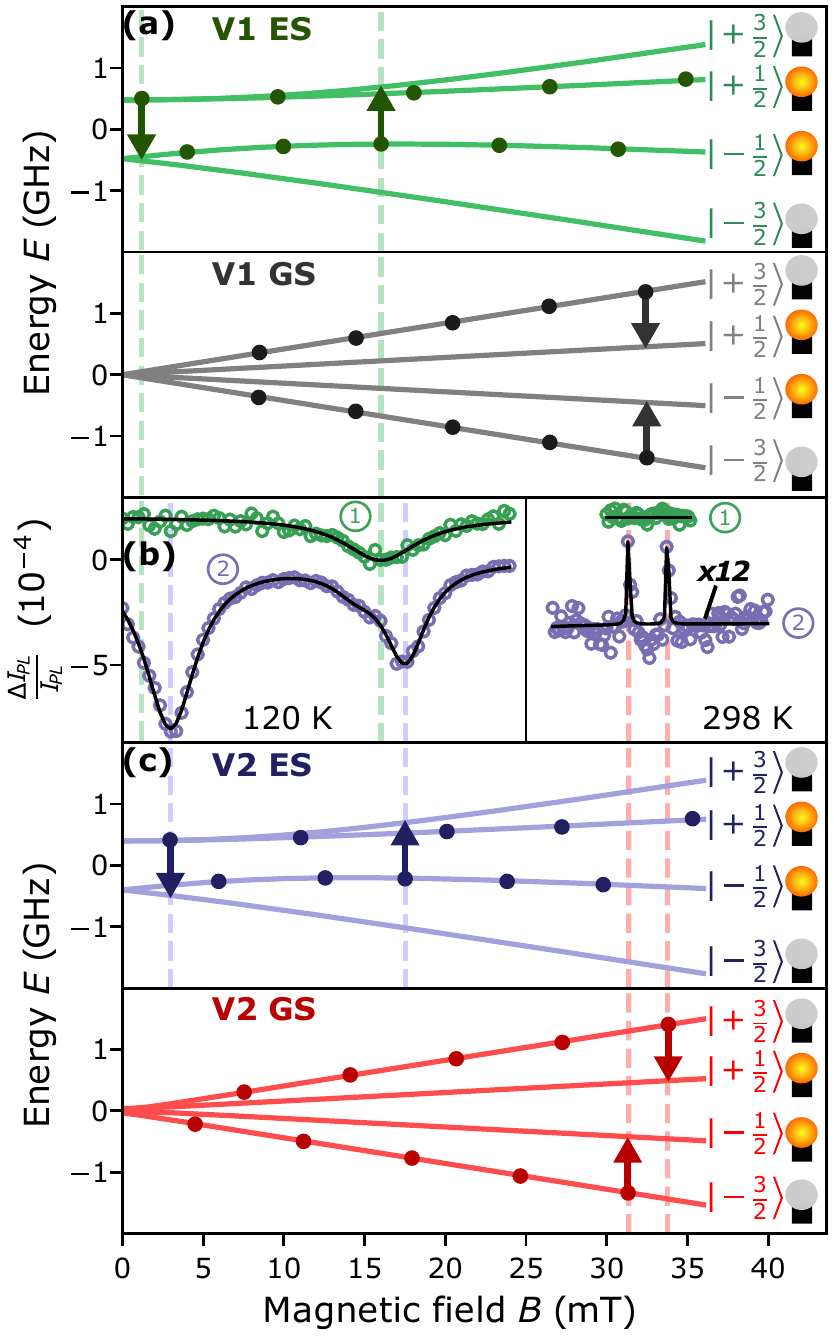}
\caption{\label{fig:overview}\textbf{(a)} Expected magnetic field dependencies of the GS and ES spin sublevels for the $V1$ center. The grey and green dots indicate the preferentially populated states under optical excitation. The PL intensity is stronger for optical transitions between the ES and GS in the $\mS=\pm1/2$ spin states, see on/off bulbs next to each spin sublevel. The green dashed line indicates the magnetic field at which SAW-induced spin resonances are observed. \textbf{(b)} ODMR measurements as a function of magnetic field for SAW induced $\Dm=\pm2$ ($B<25$~mT) and MW induced $\Dm=\pm1$ spin transitions ($B>25$~mT). The green and purple circles are measurements collecting photons in the spectral regions 1 and 2, respectively. The black curves are a multi-peak fitting of the experimental data using Lorentzian functions. \textbf{(c)} Expected magnetic field dependencies of the GS and ES spin sublevels for the $V2$ center. The purple and red dashed lines indicate the magnetic fields at which spin resonances with $\Dm=\pm2$ and $\Dm=\pm1$ are observed, respectively.}
\end{figure}

Figures~\ref{fig:overview}(a) and (c) show the energy dependence of the spin sublevels on the in-plane magnetic field for both the $V1$ and $V2$ centers, respectively. Under $B=0$, the spin sublevels in the GS and ES are split into two Kramer's doublets with energy separation $2D^{(GS,ES)}$. For $g\mu_B B \lesssim 2D$, $\mathcal{H}_{0}$ does not commute with $S_z$ and, therefore, the eigenstates of the spin Hamiltonian consist of linear combinations of the eigenstates of $S_z$, see Appendix~\ref{app_theory}. Under $g\mu_B B \gg 2D$, the Zeeman term splits all spin sublevels, and their eigenstates are well represented by those of $S_z$. For simplicity, we use this representation at strong magnetic fields to label the spin sublevels.

The selection rules for optical transitions between the GS and ES multiplets are spin conserving. However, optical excitation followed by non-radiative spin-selective relaxation via metastable states lead to a preferential population of the $\mS=\pm1/2$ states in the ES (green and purple dots in Fig.~\ref{fig:overview}) and the $\mS=\pm3/2$ states in the GS (grey and red dots in Fig.~\ref{fig:overview}), as well as a stronger PL intensity for optical transitions between the GS and ES with $\mS=\pm1/2$~\cite{HernandezMinguez:2021sa}, see light bulbs in Figs.~\ref{fig:overview}. This is in contrast to the case of an out-of-plane magnetic field, where the luminescence intensity is stronger for optical transitions between the ES and GS spin sublevels with $\mS=\pm3/2$, see Ref.~\onlinecite{Tarasenko:2017ky}.

SAW-driven spin resonances are determined by the interaction Hamiltonian~\cite{Poshakinskiy:2019bi, HernandezMinguez:2020kv}:

\begin{equation}\label{eq:SAWcoupling}
\mathcal{H}'=\Xi^{(GS,ES)}\left(\uxx S_x S_x + \uyy S_y S_y + 2 \uxy S_x S_y \right),
\end{equation}

\noindent where $\Xi^{(GS,ES)}$ is the coupling constant, which is larger for the ES than for the GS, and $\uij$ are the components of the strain tensor. For a SAW propagating along the $x$ direction, the only non-zero strain components are the in-plane and out-of-plane longitudinal strains, $u_{xx}$ and $u_{yy}$, respectively, as well as the shear strain $u_{xy}$. When the in-plane magnetic field is applied perpendicularly to the $x$ axis, the transition rates for SAW-induced spin resonances with $\Dm=\pm2$ are maximized, but are suppressed for the $\Dm=\pm1$ ones~\cite{HernandezMinguez:2020kv}, even in the presence of spin mixing caused by the ZFS term in Eq.~\ref{eq:SpinHamiltonian}, see Appendix~\ref{app_selrules}.

\section{Results}\label{sec:Results}

Figure~\ref{fig:overview}(b) compares aODMR measurements taken in the spectral regions of the $V1$ and $V2$ centers (green and purple circles, respectively). In agreement with our previous results~\cite{HernandezMinguez:2020kv, HernandezMinguez:2021sa}, the strain field of the SAW drives two broad resonances at low magnetic fields (below 25~mT) corresponding to the $\Dm=\pm2$ transitions between the ES spin sublevels of the $V2$ center, see purple vertical arrows in Fig.~\ref{fig:overview}(c). At large magnetic fields (above 25~mT), the $\Dm=\pm1$ transitions are not acoustically excited, and we observe only two weak ODMR resonances (note that the ODMR signal is amplified more than 10 times), which originate from the weak MW stray field emitted by the IDT, see Appendix~\ref{app_selrules}. This is confirmed by the fact that ODMR measurements taken with the acoustic resonator excited at MW frequencies out of the IDT emission band show only these two narrow peaks corresponding to the $\Dm=\pm1$ spin transitions between the GS sublevels of the $V2$ center~\cite{HernandezMinguez:2021sa}, see the red vertical arrows in Fig.~\ref{fig:overview}(c).

\begin{figure}
\includegraphics[width=\columnwidth]{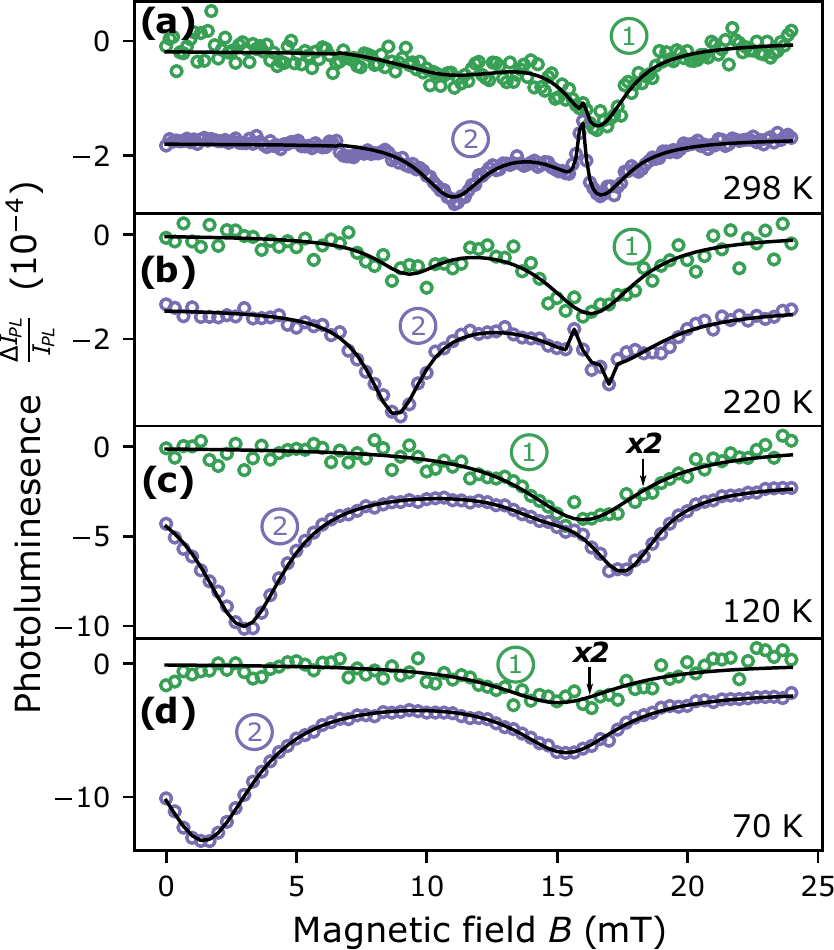}
\caption{\label{fig:ODMRvsT} aODMR measurements at \textbf{(a)} 298 K, \textbf{(b)} 220 K, \textbf{(c)} 120 K and \textbf{(d)} 70 K, obtained by collecting the PL emitted in the spectral region 1 (green circles) and 2 (purple circles). The black curves are multi-peak fittings of the experimental data using Lorentzian functions. The aODMR measurements are vertically shifted for clarity.}
\end{figure}

The situation is different for the measurements taken at the spectral range of the $V1$ center. As in the case of the $V2$ center, we do not detect any aODMR signal at the magnetic field region of the $\Dm=\pm1$ spin transitions (see grey vertical arrows in Fig.~\ref{fig:overview}(a)). However, at low magnetic fields, we observe a single broad dip at a magnetic field slightly lower than the one for the $V2$ center. To better understand the nature of this spin resonance, we repeated the measurement at several temperatures and fitted the experimental data with Lorentzian functions to obtain the resonant magnetic fields. Figure \ref{fig:ODMRvsT} summarizes the results for four different temperatures. In all cases, the aODMR measurement in the spectral region of the $V1$ center (green circles) shows the single broad dip with an approximately temperature independent amplitude, but at a magnetic field that shifts from 17~mT to 15~mT as the temperature drops. In contrast, the number of peaks and dips observed in the spectral region of the $V2$ centers (purple circles) depends on temperature. Above 200~K, we observe the broad resonances of the ES together with narrow resonances around 16 mT corresponding to the $\Dm=\pm2$ GS spin transitions, as have been described in our earlier work~\cite{HernandezMinguez:2021sa}. In addition, the broad dip attributed to the $V1$ center is partially observed due to the fact that a fraction of the photons from its Stokes PSB are emitted within the spectral region of the $V2$ center. Conversely, photons from the anti-Stokes PSB of the $V2$ center make the aODMR measurements in the spectral region of the $V1$ center partially sensitive to the $V2$ resonances (see weak dips around 10~mT for 298~K and 220~K). As the temperature decreases, the reduction in the PSB emission suppresses this cross-detection effect. In addition, below 200~K the narrow GS spin transitions of the $V2$ center cannot be resolved due to the large amplitude of the ES spin resonances in this temperature range.

\begin{figure}
\includegraphics[width=\columnwidth]{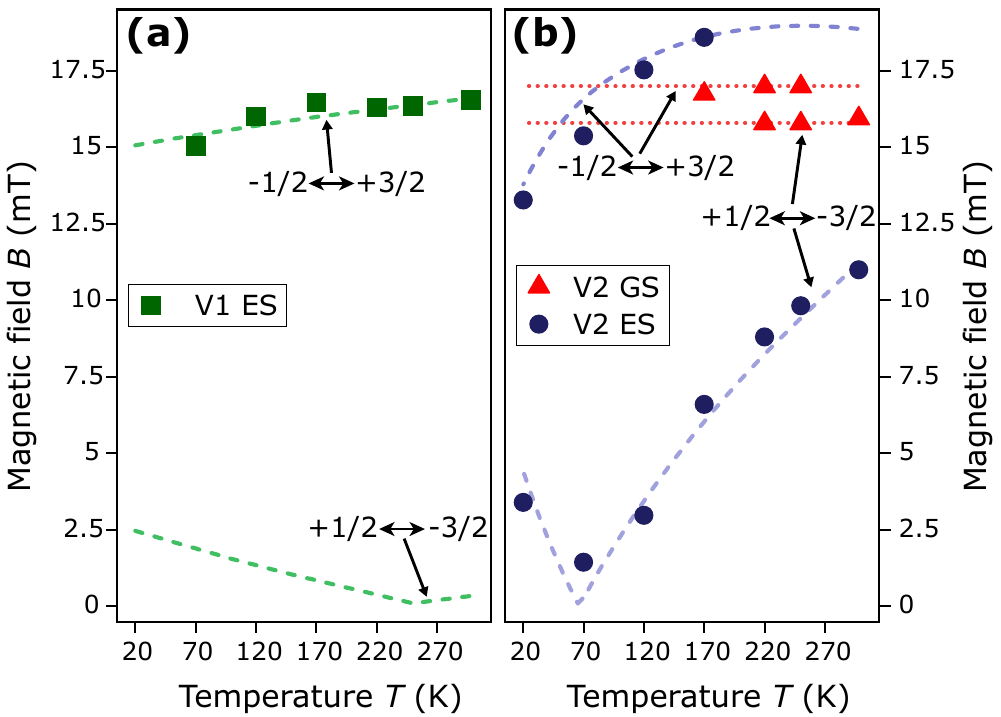}
\caption{\label{fig:Tdep} Magnetic fields of the spin resonances as a function of temperature for (a) the $V1$ center and (b) the $V2$ center. The solid symbols are the measured magnetic fields, while the dotted and dashed lines are the theoretically calculated temperature dependencies for the ground and excited states, respectively. The size of the symbols is larger than the error of the measured magnetic fields.}
\end{figure}

Figure~\ref{fig:Tdep} summarizes the magnetic fields of all identified spin resonances for all measured temperatures. The red triangles and purple circles in Fig.~\ref{fig:Tdep}(b) are the results of the Lorentzian fits in Fig.~\ref{fig:ODMRvsT} for the GS and ES spin resonances of the $V2$ centers, respectively. The red dotted and purple dashed curves represent the theoretical behavior calculated using Eq.~\ref{eq:SpinHamiltonian} and taking into account the temperature dependencies of $2D^{(GS)}$ and $2D^{(ES)}$ reported in Ref.~\onlinecite{Anisimov:2016er}. While the splitting in the GS is nearly temperature independent with $2D^{(GS)}/h=70$ MHz, it depends linearly on the temperature in the ES according to the equation $2D^{(ES)}/h=1060~\mathrm{MHz}-2.1~\mathrm{MHz/K} \cdot T$. This is justified by the fact that the typically larger extension of the ES electronic wave function makes $2D^{(ES)}$ much more sensitive to temperature-induced changes in the crystallographic environment than $2D^{(GS)}$. Therefore, for a fixed SAW frequency, the magnetic fields of the GS resonances are independent of temperature, while the ES resonances move to lower magnetic fields as the temperature decreases. Note that the temperature dependence of the ES resonances is not linear due to the different orientations of the spin operator in the Zeeman and ZFS terms of the spin Hamiltonian, see Eq.~\ref{eq:SpinHamiltonian}.

\section{Discussion}\label{sec:Discussion}

We now discuss the origin of the spin resonances attributed to the $V1$ center, see green squares in Fig.~\ref{fig:Tdep}(a). In contrast to $V2$, the splitting amplitudes of $V1$ have so far only been measured at 4~K with values $2D^{(GS)}/h=4$~MHz and $2D^{(ES)}/h=985$~MHz~\cite{Nagy:2019fw}. Therefore, we will assume here that the mechanisms leading to the temperature dependencies of the ZFS parameters for the $V2$ center are also valid for the $V1$ center. Assuming a temperature independent $2D^{(GS)}$, we obtain that $2D^{(GS)}/h\ll\fSAW$ and the ZFS contribution to the GS spin Hamiltonian can be neglected for all temperatures. Then, both $\Dm=\pm2$ GS spin transition frequencies will equal the SAW frequency at $\B=h\fSAW/\vert g\mu_B\Dm\vert=16.5$~mT. This value agrees well with the magnetic field of the spin resonance attributed to the $V1$ center at room temperature, thus suggesting that it may correspond to a GS spin transition. However, GS spin resonances are typically characterized by narrow line widths and ES spin resonances by wide ones, due to the fact that the spin coherence time in the ES is limited by its short optical relaxation time. In addition, the amplitude and width of the spin resonance attributed to the $V1$ center are comparable to the ES resonances of the $V2$ center, thus suggesting that the observed $V1$ spin transitions should rather be attributed to the ES rather than the GS. Finally, the $V1$ resonance shifts to lower magnetic fields at low temperatures, following a behavior similar to the ES $-1/2 \rightarrow +3/2$ spin resonance of the $V2$ center, although with a weaker temperature dependence.

To study this possibility in more detail, we have estimated the temperature dependence of $2D^{(ES)}$ for the $V1$ center in the following way. First, we have considered the previously reported value of $2D^{(ES)}/h=985$ MHz at 4 K~\cite{Nagy:2019fw}. Then, we have calculated the value of $2D^{(ES)}$ for which the magnetic field of the ES $-1/2 \rightarrow +3/2$ spin resonance coincides with the measured value at room temperature, obtaining $2D^{(ES)}/h=910$ MHz. A linear fitting between these two values leads to the equation:

\begin{equation}\label{eq:DvsT}
2D^{(ES)}/h \;= \;986.28~\mathrm{MHz}\;-\;0.254~\mathrm{MHz/K}\;\cdot\;T.
\end{equation}

\noindent The dashed green curves in Fig.~\ref{fig:Tdep}(a) show the magnetic fields at which both $\Dm=\pm2$ ES spin resonances take place for the full temperature range, calculated using Eqs.~\ref{eq:SpinHamiltonian} and \ref{eq:DvsT}. The experimental data agree well with the theoretical values for the ES $-1/2 \rightarrow +3/2$ spin transition. Taking into account the similar amplitude and width of the aODMR for the two spin centers, the weak temperature dependence of $2D^{(ES)}$ for the $V1$ center (about eight times weaker than for the $V2$ center) is a remarkable difference.

Finally, the theory also predicts an ES spin resonance at low magnetic fields (dashed green curve below 3 mT in Figs.~\ref{fig:overview}(a) and \ref{fig:Tdep}(a)), which we did not detect in our aODMR experiments. It has been reported that the amplitude and sign of the spin resonances in the ES depend on the experimental conditions~\cite{Hoang:2021dx}. In our case, the different degree of spin mixing of the GS, ES and metastable states caused by the in-plane magnetic field, and the fact that the second ES of the $V1$ center is energetically close to the first ES, may lead to different rates for the optical and spin relaxation processes than for the $V2$ center, thus making the low magnetic field resonance of the first ES not optically addressable for certain temperature and magnetic field ranges. A detailed understanding of all these phenomena would require additional studies that go beyond the scope of this manuscript.
 
\section{Conclusions and outlook}\label{sec:Conclusion}
 
In summary, we have demonstrated acoustically induced spin transitions both in the $V1$ and $V2$ centers in 4H-SiC. Based on the width and temperature dependence of the spin resonance attributed to the $V1$ center, we assign it to a transition between spin sublevels in the excited state. In contrast to MW-induced spin resonances, SAW-induced spin resonances are observed for both the $V1$ and $V2$ centers at all temperatures studied, up to room temperature, thus suggesting that both kinds of centers have similar  sensitivity to the dynamic strain of the SAW. However, the number of observed resonances and their temperature dependencies are significantly different for the two centers. This is attributed to the different impact of their non-common local crystallographic environments on the orbital and spin energy levels, as well as on the relaxation rates between the ground, excited and metastable states. Additional measurements using high-frequency SAW resonators (above 3 GHz) and magnetic fields (above 50 mT), where the effects of the crystallographic environment on the spin sublevels can be neglected, should contribute to a better understanding of the different response of the $V1$ center to the acoustic excitation compared to its $V2$ counterpart.

The acoustic spin control of both $V1$ and $V2$ centers in their excited states opens new possibilities for e.g. the implementation of efficient quantum sensing protocols using spin-optomechanics~\cite{Soykal:2016tk, Poshakinskiy:2019bi}. The larger thermal response of the zero-field splitting in the excited state, together with the stronger ODMR contrast of the acoustically driven spin resonances, should lead to a thermal sensitivity up to two orders of magnitude larger than for the ground state, which, in turn, is already one order of magnitude larger than for NV centers in diamond~\cite{Acosta:2010fq}. Here, potential limitations caused by the typically broad ODMR resonances in the excited state could be overcome by taking advantage of the previously reported coherent spin trapping mechanism~\cite{HernandezMinguez:2021sa}, where the simultaneous acoustic excitation of the same spin transition both in the GS and ES leads to a strong sensitivity of the ODMR signal in the ground state to the thermal shift of the spin resonance in the excited state. In addition, the simultaneous acoustic control of the $V1$ and $V2$ centers could be used to e.g. implement calibration-free sensing schemes combining different magnitudes like temperature and magnetic field.

Regarding quantum spin control, the strong sensitivity of the excited states to acoustic vibrations could allow for new and fast methods of quantum information processing. As an example, a SAW beam tuned to a certain spin transition in the excited state can manipulate the quantum information stored in the spin center only during the short time that it stays in the excited state. By using selective optical excitation to address a particular spin center, it should be possible to independently control many spin qubits integrated in a single acoustic resonator. Finally, due to the large extension of the electronic wave function in the excited state, the efficient acoustic control of its spin multiplet opens promising ways for the efficient manipulation of nearby nuclear spins~\cite{Jacques:2009fo, Falk:2015iz, Ivady:2015kd}, a key ingredient for applications in quantum technologies.

\begin{acknowledgments}
The authors would like to thank S. Meister and S. Rauwerdink for technical support in the preparation of the sample, and O. Brandt for a critical reading of the manuscript. G. V. A. acknowledges the support from the German Research Foundation (DFG) under Grant No. AS 310/9-1. The authors acknowledge support from the Ion Beam Center (IBC) at Helmholtz-Zentrum Dresden-Rossendorf (HZDR) for the proton irradiation.
\end{acknowledgments}

\appendix

\section{Theoretical description of spin system}\label{app_theory}

Diagonalization of Eq.~\ref{eq:SpinHamiltonian} gives the energy $E_\alpha$ of the four spin eigenstates as a function of magnetic field strength:

\begin{eqnarray}
E_{+3/2} &=& \frac{\gamma B}{2} + \sqrt{D^2 - \gamma B D + (\gamma B)^2} ,\\
E_{+1/2} &=& -\frac{\gamma B}{2} + \sqrt{D^2 + \gamma B D + (\gamma B)^2} ,\\
E_{-1/2} &=& \frac{\gamma B}{2} - \sqrt{D^2 - \gamma B D + (\gamma B)^2} ,\\
E_{-3/2} &=& -\frac{\gamma B}{2} - \sqrt{D^2 + \gamma B D + (\gamma B)^2},
\end{eqnarray}

\noindent where $\gamma=g\mu_B$, and the subscript $\alpha$ denotes the projection of the spin operator along the $z$ direction when $D=0$.

Under $D \neq 0$, $\mathcal{H}_{0}$ does not commute with $S_z$ and the eigenstates of the Hamiltonian, $\ket{E_\alpha}$, do not coincide with those of $S_z$. Taking into account that $S_x$ and $S_y$ can be rewritten as $S_x=\frac{1}{2}(S_+ + S_-)$ and $S_y=\frac{1}{2i}(S_+ - S_-)$, where $S_+$ and $S_-$ are the raising and lowering operators, respectively, the eigenstates of Eq.~\ref{eq:SpinHamiltonian} are expressed, up to a normalization factor, as the following linear combinations of the eigenstates of $S_z$:

\begin{eqnarray}
\ket{E_{+3/2}} &=& \ket{+3/2}-a(D,B)\ket{-1/2},\label{eq:eigenstate1}\\
\ket{E_{+1/2}} &=& \ket{+1/2}-b(D,B)\ket{-3/2},\label{eq:eigenstate2}\\
\ket{E_{-1/2}} &=& \ket{-1/2}+a(D,B)\ket{+3/2},\label{eq:eigenstate3}\\
\ket{E_{-3/2}} &=& \ket{-3/2}+b(D,B)\ket{+1/2}.\label{eq:eigenstate4}
\end{eqnarray}

\noindent Here, the coefficients $a,b \rightarrow 0$ when $2D/\gamma B\rightarrow 0$.

\section{Selection rules of spin transitions}\label{app_selrules}

\subsection{SAW-induced spin transitions}

The effective Hamiltonian that couples the spin operator and the components of the strain tensor, $\uij$, is~\cite{Udvarhelyi:2018bx, Udvarhelyi:2018tg}:

\begin{equation}\label{eq:Hsp1}
\mathcal{H}' =\sum_{ijkl}\Xi_{ijkl}\uij S_k S_l,
\end{equation}

\noindent where $\Xi_{ijkl}$ is the 4th-order range deformation potential tensor, which in the case of the spherical approximation can be reduced to a constant. For a SAW propagating along the $x$ direction, the only non-zero strain components are the in-plane and out-of-plane longitudinal strains, $\uxx$ and $\uyy$, respectively, and the shear strain $\uxy$. Therefore, Eq.~\ref{eq:Hsp1} simplifies to:

\begin{equation}\label{eq:Hsp2}
\mathcal{H}' =\Xi\left( \uxx S_x S_x + \uyy S_y S_y + 2\uxy S_x S_y \right).
\end{equation}

\noindent By expressing the operators $S_x$, $S_y$ as a function of $S_+$ and $S_-$, Eq.~\ref{eq:Hsp2} can be rewritten as a linear combination of the product operators $S_+S_+$, $S_+S_-$, $S_-S_+$ and $S_-S_-$. Therefore, when $\mathcal{H}'$ is applied to the eigenstates of $S_z$, the only non-zero transition amplitudes $\bra{m'}\mathcal{H}'\ket{m}$ are those fulfilling the conditions $m'=m$ (due to the $S_+S_-$ and $S_-S_+$ operators) or $m'=m\pm2$ (due to $S_+S_+$ and $S_-S_-$). Taking these rules into account and the representations of $\ket{E_\alpha}$ as linear combinations of the eigenstates of $S_z$, the only SAW-induced spin transitions with non-zero probabilities are those fulfilling $\alpha'-\alpha=\pm2$, that is $\vert\bra{E_{-1/2}}\mathcal{H}'\ket{E_{+3/2}}\vert^2$ and $\vert\bra{E_{+1/2}}\mathcal{H}'\ket{E_{-3/2}}\vert^2$. This argument is valid for all values of $2D/\gamma B$.

\subsection{MW-induced spin transitions}

In the presence of a stray MW field, the Hamiltonian coupling $\vS$ and the oscillating magnetic field, $\vbMW=(b_x,b_y,b_z)$, is:

\begin{equation}
\mathcal{H}'= g\mu_B\left(b_xS_x+b_yS_y+b_zS_z\right).
\end{equation}

\noindent This coupling Hamiltonian is linear in the components of the spin operator. By rewriting $S_x$ and $S_y$ in terms of $S_+$ and $S_-$, that the non-zero transition amplitudes $\bra{m'}\mathcal{H}'\ket{m}$ between the four eigenstates of $S_z$ are those fulfilling the conditions $m'=m$ (due to the $S_z$ operator) and $m'=m\pm1$ (due to $S_+$ and $S_-$). By applying these selection rules to the calculation of $\vert\bra{E_{\alpha'}}\mathcal{H}'\ket{E_{\alpha}}\vert^2$, it comes out that, in principle, all possible transitions between the four eigenstates of $\mathcal{H}_{0}$ are allowed for the case of $D\neq0$ and a MW stray field with arbitrary $\vbMW$ direction. 

However, for the ground state of the $\VSi$ center, $2D/\gamma B\ll 1$ for $B\gg 2.5$~mT. Since all the spin transitions between the $\ket{E_{\alpha}}$ eigenstates match the SAW frequency at magnetic fields much larger than 2.5~mT, the contribution of $D$ to $\mathcal{H}_{0}$ can be neglected. Therefore, the stray MW field of the IDT only excites transitions between the $\ket{E_\alpha}$ eigenstates that fulfill the condition $\alpha'-\alpha=\pm1$.

In the case of the spin resonances in the excited state observed in Figs. 2 and 3 of the manuscript, the magnetic fields at which they take place are much lower that the ones required to fulfill the condition $2D/\gamma B\ll 1$. Therefore, such transitions could be driven by the stray MW field of the IDT. However, in our experiment, the stray MW field is not strong enough to drive spin transitions during the short time that the color center stays in the excited state. We have confirmed this in our previous work, see Ref.~21, where we recorded the ODMR signal under a MW frequency out of the resonance of the IDT, and therefore in the absence of SAWs. The only changes observed in the PL intensity appeared at the magnetic fields where the MW frequency matched the $\alpha'-\alpha=\pm1$ transitions of the ground state.



%

\end{document}